\documentclass[prl,twocolumn,showpacs]{revtex4}
\usepackage{amsmath,amsthm,amssymb,amscd,float,accents}
\usepackage{graphicx}
\usepackage[T1]{fontenc}
\usepackage{lmodern}
\newcommand{\al}{\alpha}
\newcommand{\be}{\beta}
\newcommand{\de}{\delta}
\newcommand{\ep}{\epsilon}

\newcommand{\si}{\sigma}

\newcommand{\vp}{\varphi}

\newcommand{\bs}{\mathbf{s}}

\newcommand{\cE}{{\mathcal E}}

\newcommand{\cH}{{\mathcal H}}

\newcommand{\cM}{{\mathcal M}}

\newcommand{\pa}{\partial}

\def\ket#1{|#1\rangle}

\renewcommand{\leq}{\leqslant}
\renewcommand{\geq}{\geqslant}
\renewcommand{\le}{\leqslant}
\renewcommand{\ge}{\geqslant}
\def\bbuildrel#1_#2^#3{\mathrel{\mathop{\kern0pt #1}\limits_{#2}^{#3}}}
\newcommand{\tends}[1]{\bbuildrel{\hbox to 2em{\rightarrowfill}}_{#1}^{}}
\newcommand{\iu}{\mathrm{i}}
\newcommand{\e}{\mathrm{e}}

\newcounter{ex}

\begin{document}
\title{Spin chains of Haldane--Shastry type and a generalized central limit theorem}
\author{Alberto \surname{Enciso}}%
\email{alberto.enciso@math.ethz.ch} \affiliation{Departement Mathematik, ETH Zürich, 8092 Zürich,
  Switzerland} \author{Federico \surname{Finkel}}%
\email{ffinkel@fis.ucm.es} \author{Artemio \surname{González-López}}%
\email[Corresponding author. Electronic address: ]{artemio@fis.ucm.es}%
\affiliation{Departamento de Física Teórica II, Universidad Complutense, 28040 Madrid, Spain}
\date{March 26, 2009}
\begin{abstract}
  We show that the density of energy levels of a wide class of finite-dimensional
  quantum systems tends to a Gaussian distribution as the number of degrees of freedom increases.
  Our result is based on a nontrivial modification of the classical central limit theorem, and is
  especially suited to models whose partition function is explicitly known. In particular, we
  provide the first theoretical explanation of the fact that the level density of several spin
  chains of Haldane--Shastry type is asymptotically Gaussian when the number of sites tends to
  infinity.
\end{abstract}
\maketitle
%

Spin chains of Haldane--Shastry (HS) type~\cite{Ha88,Sh88} are the prime example of integrable
spin chains with long-range interactions, having close connections with several topics of current
interest such as strongly correlated systems~\cite{ASK01}, generalized exclusion
statistics~\cite{MS94,Po06}, and the AdS-CFT correspondence~\cite{HL04}. Recent numerical
computations have shown that a common feature of these chains~\cite{EFGR05,FG05,BFGR08,BFGR08epl}
and their supersymmetric extensions~\cite{BB06,BFG09,BB09} is the fact that the level density
(normalized to $1$) becomes Gaussian when the number of sites $N$ tends to infinity. The knowledge
of the continuous part of the level density is a key ingredient in the theory of quantum chaos,
since it is used to rescale (``unfold'') the spectrum as a preliminary step in the study of
important statistical properties such as the distribution of spacings between consecutive levels
\cite{Ha01}. In fact, a long-standing conjecture of Berry and Tabor~\cite{BT77} posits that the
spacings distribution of a ``generic'' integrable system should be Poissonian, while in quantum
chaotic systems like polygonal billiards this distribution is given by Wigner's
surmise~\cite{Me04}, characteristic of the Gaussian ensembles in random matrix theory. For many
spin chains of HS type, it can be shown that the Gaussian character of the level density implies
that the spacings distribution obeys neither Poisson's nor Wigner's law, but is rather given by a
simple ``square-root of a logarithm'' formula~\cite{BFGR08,BFGR08epl,BFG09,BB09}. In this letter
we develop a generalization of the standard central limit theorem to show that the level density
of a wide class of finite-dimensional quantum systems must be asymptotically Gaussian. This class
includes in particular a supersymmetric version of the original (trigonometric) Haldane--Shastry
spin chain, as well the $BC_N$ version~\cite{BFGR08} of the spin $1/2$ Polychronakos--Frahm
(rational) chain~\cite{Po93,Fr93}.

To explain the gist of our approach, let us begin by considering an extremely simple toy model for
which the asymptotically Gaussian character of the spectrum can be easily established. Indeed, let
$\cH$ be a finite-dimensional Hilbert space and let us denote by $\cH^{\otimes N}$ its $N$-th tensor
product, which can be identified with the $N$-particle space. Our toy Hamiltonian will be given by
\begin{equation}\label{toy}
  H=I_1+\cdots+ I_N\,,
\end{equation}
where each operator
$I_k=1_{\cH}\otimes\cdots\otimes\smash{\overset{\,\underaccent{\smile}{k}}{I}}\otimes\cdots\otimes
1_{\cH}$ is a constant of motion acting on the Hilbert space of the $k$-th particle as some fixed
operator $I$. Physically, this system can be thought of as consisting of $N$ identical,
non-interacting particles. Had we allowed the one-particle Hilbert space to be infinite
dimensional, this would precisely be the structure underlying the $N$-dimensional harmonic
oscillator.

It is not difficult to see that in the limit $N\to\infty$ the level density of the
Hamiltonian~\eqref{toy} becomes Gaussian, with mean $\mu=N\mu_I$ and variance $\si^2=N\sigma_I^2$
($\mu_I$ and $\si_I^2$ respectively denoting the mean and variance of the spectrum of $I$).
Indeed, the eigenvalues of $H$ are given by
\begin{equation}\label{E}
  \cE=\cE_1+\cdots+\cE_N\,,
\end{equation}
where each $\cE_k$ is an eigenvalue of the operator $I$. Since each of these eigenvalues can be
freely chosen, the parameters $\cE_k$ in the above formula can be regarded as independent random
variables with the same distribution. Thus, the fact that the random variable $\cE$ asymptotically
follows the Gaussian law as $N\to\infty$ is an immediate consequence of the central limit theorem.

It is natural to wonder if the simple argument above can be extended to a wider class of models. A
cursory inspection reveals, however, that several assumptions must be substantially relaxed in
order to cover any physically interesting situation. In particular, in Eq.~\eqref{E} it is crucial
to allow for sums of independent random variables which are {\em not}\/ identically distributed.
It is also clear that it would be preferable to express the conditions on the spectrum of our
model in terms of its partition function rather than its eigenvalues. Indeed, for chains of HS
type the partition function can be computed in closed form through Polychronakos's ``freezing
trick''~\cite{Po93,Po94}, while the spectrum is considerably more difficult to handle. Thus, in
order to further develop the former approach, we will need to prove a
modification of the classical central limit theorem incorporating the above requirements.

Let us introduce some notation before going on. We shall denote by
\[
Z(q)=\sum_{i=1}^nd_iq^{\cE_i}
\]
the partition function of a finite-dimensional Hamiltonian $H$ with energy levels $\cE_i$ and
degeneracies $d_i$, $1\le i\le n$, where $q=\e^{-1/(k_{\rm B}T)}$ and $k_{\rm B}$ is Boltzmann's
constant. The value $Z(1)=\sum_id_i$ is thus the dimension of the Hilbert space of the system. Its
thermodynamic energy at temperature $T$ is given by
\[
E(q)=q\,\frac{\pa}{\pa q}\log Z(q)\,,
\]
in terms of which the mean and variance of the spectrum of $H$ can be expressed as
\begin{align*}
  \mu&=\langle H\rangle\equiv\frac1{Z(1)}{\sum_{i=1}^nd_i\,\cE_i}=E(1)\,,\\
  \si^2&=\big\langle(H-\mu)^2\big\rangle
  \equiv\frac1{Z(1)}{\sum_{i=1}^nd_i\,\big(\cE_i-\mu\big)^2}=E'(1)\,.
\end{align*}

We will henceforth restrict our attention to systems whose partition function can be written as a
product
\begin{equation}\label{Z}
  Z(q)=\prod_{k=1}^NZ_k(q)\,,
\end{equation}
and we shall assume that the factors $Z_k$ possibly depend on the integer parameter $N$ in view of
forthcoming applications. Roughly speaking, the system is equivalent to an effective model of $N$
non-interacting, but not necessarily identical, subsystems. Two powerful tools in the study of the
energy distribution of~\eqref{Z} are the characteristic function
\begin{equation}\label{vp}
  \vp(t)\equiv\big\langle\e^{\iu t H}\big\rangle=\frac{Z(\e^{\iu t})}{Z(1)}\,,
\end{equation}
which is essentially its Fourier transform, and the normalized characteristic
function
\begin{equation}\label{barvp}
  \bar\vp(t)\equiv\Big\langle\e^{\iu t \frac{H-\mu}\si}\Big\rangle
  =\e^{-\frac{\iu t\mu}\si}\vp\big(\tfrac t\si\big)\,.
\end{equation}
The key property of the latter function is that, under very mild technical
assumptions~\cite{AD00}, in the limit $N\to\infty$ the energy distribution of the system
asymptotically follows the Gaussian law with mean $\mu$ and standard deviation $\si$ if and only
if
\begin{equation}\label{lim}
  \lim_{N\to\infty}\bar\vp(t)=\e^{-\frac12 t^2}\,.
\end{equation}

We shall next provide some simple conditions on the factors $Z_k$ ensuring that Eq.~\eqref{lim}
holds. In order to state them in a concise way, we will use a subscript $k$ (as in $\mu_k$ or
$\bar\vp_k$) to denote the statistical quantities computed with respect to the $k$-th partition
function $Z_k$, and introduce the notation
\[
\cM_k(\tau)\equiv\sup_{|s|<\tau}\bigg|\frac{\pa^3}{\pa s^3}\log \bar\vp_k(s)\bigg|\,.
\]
It should be noticed that the average and standard deviation of the energy of the whole system are
related to the analogous quantities of its subsystems by
\begin{equation}\label{relation}
  \mu=\sum_k\mu_k\,,\qquad \si^2=\sum_k\si_k^2\,,
\end{equation}
where the index runs from $1$ to $N$ (as always hereafter, unless otherwise stated). Our
assumptions on the factors are then the following:
\begin{enumerate}
\item The standard deviation of the full partition function and of its $k$-th factor satisfy
  \begin{equation}\label{cond.sigma}
    \frac{\si_k}\si\leq C_1N^{-\frac12}\,,\qquad  1\le k\le N\,,
  \end{equation}
  where the constant $C_1$ does not depend on $N$.
\item There exist some positive constants $\ep_1,\ep_2,C_2$, independent of $N$, such that
  \begin{equation}\label{cM}
    \cM_k(\ep_1)\leq C_2 N^{\frac12-\ep_2}\,,\qquad  1\le k\le N\,.
  \end{equation}
\end{enumerate}
Let us briefly discuss the meaning of these assumptions. Roughly speaking, the first condition
holds whenever the standard deviation of the whole system does not effectively depend only on a
few subsystems. More precisely, let
\[
M(N)\equiv\max\big\{\si_k^2:k=1,\dots,N\big\}\,,
\]
and suppose that there exists a constant $\al>0$ (independent of $N$) and a function $m(N)>0$ such
that
\[
\si_k^2\ge m(N)
\]
for at least $\al N$ values of $k$. Then, a sufficient condition ensuring the validity of (i) is
that there exists a constant $A$ such that
\begin{equation}\label{criterion}
  M(N)\leq A\, m(N)
\end{equation}
as $N\to\infty$. Indeed, by definition of $m(N)$ and Eq.~\eqref{relation} we have that
$\si^2\geq\al N\cdot m(N)$, which in turn implies that
\[
\frac{\si_k}{\si}\leq\bigg(\frac{M(N)}{\al N m(N)}\bigg)^{\frac12}\leq C_1N^{-\frac12}\,,
\qquad 1\leq k\leq N\,,
\]
with $C_1=(A/\al)^{1/2}$. On the other hand, the technical condition~(ii) is a local bound on the
third central moment of the thermodynamic energy of the $k$-th subsystem at imaginary temperature.
Indeed, it can be shown that
\begin{multline*}
  \frac1{\iu^3}\,\frac{\pa}{\pa s^3}\log\bar\vp_k(s)=
\frac1{\iu^3}\,\frac{\pa}{\pa s^3}\log Z_k\big(\e^{\frac{\iu s}{\si_k}}\big)\\
  =\bigg(\frac{q}{\si_k}\frac{\pa}{\pa_q}\bigg)^3\log Z_k(q)\Big|_{q=\e^{\frac{\iu s}{\si_k}}}
  =\frac1{\si_k^3}\,\overline{\big(\cE_k-E_k(q)\big)^3}\,\Big|_{q=\e^{\frac{\iu s}{\si_k}}}\,,
\end{multline*}
where the overbar denotes thermal average.

It should also be noted that our second condition is reminiscent of Lyapunov's classical
condition~\cite{AD00}, according to which~\eqref{lim} holds if there is a positive $\de$ such that
\begin{equation}\label{lyapunov}
\lim_{N\to\infty}\frac1{\si^{2+\de}}\sum_k\big\langle|\cE_k-\mu_k|^{2+\de}\big\rangle=0\,.
\end{equation}
Note, however, that due to the absolute value the latter condition cannot be expressed in terms of
the partition functions $Z_k$ (this is also true for the more general condition due to
Lindeberg~\cite{AD00}). Thus~\eqref{lyapunov} is impractical when the partition function is known
but there is not an explicit formula for the spectrum, as is the case with spin chains of HS type.

In order to see that conditions (i) and (ii) above imply that the level density asymptotically
follows the Gaussian law, let us compute the limit of $\log\bar\vp(t)$ as $N\to\infty$. To begin
with, one should observe that the fact that $\bar\vp_k$ is normalized to zero mean and unit
variance ensures that the second-order Taylor expansion of $\log\bar\vp_k$ around $0$ is
\[
\log\bar\vp_k(\tau)=-\frac{\tau^2}2+R_k(\tau)\,,
\]
where the remainder is bounded by
\begin{equation}\label{Rk}
  \big|R_k(\tau)\big|\leq\frac{\tau^3}6\cM_k(|\tau|)\,.
\end{equation}
Since $\vp(t)=\prod_k\vp_k(t)$ by Eq.~\eqref{vp}, it immediately follows from~\eqref{barvp}
and~\eqref{relation} that
\[
\log\bar\vp(t)=\sum_k\log\bar\vp_k\bigg(\frac{\si_kt}\si\bigg)=-\frac{t^2}2+\sum_k
R_k\bigg(\frac{\si_kt}\si\bigg)\,,
\]
where by virtue of Eq.~\eqref{Rk} the error can be estimated as
\begin{equation}\label{error}
  \bigg|\log\bar\vp(t)+\frac{t^2}2\bigg|
  \leq \frac16\sum_k\bigg(\frac{\si_k|t|}\si\bigg)^3 \cM_k\bigg(\frac{\si_k|t|}\si\bigg)\,.
\end{equation}
Let us assume that $|t|<t_0$, where $t_0$ is a fixed but otherwise arbitrary constant, and take
$N$ greater than $\big(\frac{C_1t_0}{\ep_1}\big)^2$. In this case, by~\eqref{cond.sigma} we have
\[
\frac{\si_k|t|}\si\leq C_1t_0N^{-\frac12}\leq\ep_1\,,
\]
so that~\eqref{error} can be controlled as
\[
\bigg|\log\bar\vp(t)+\frac{t^2}2\bigg|\leq \frac16(C_1t_0)^3C_2N^{-\ep_2}\,,
\]
on account of~\eqref{cM}. It then follows that $\log\bar\vp(t)$ converges pointwise to
$-\frac12t^2$ for all real $t$, the convergence being uniform on compact sets. Hence
Eq.~\eqref{lim} holds, and thus the level density becomes asymptotically Gaussian by the
properties of characteristic functions, as we wanted to show.

A particularly simple class of partition functions of the form~\eqref{Z} is obtained by requiring
that each factor $Z_k$ corresponds to a two-level system. In this case, setting the ground state
energy to zero without loss of generality, each factor can be written as
\begin{equation}\label{Zk}
  Z_k(q)=1+q^{\cE(k,N)}\,,
\end{equation}
and the mean and standard deviation of its energy are readily computed as
$\mu_k=\si_k=\frac12\cE(k,N)\,.$ Its reduced characteristic function is simply
$\bar\vp_k(\tau)=\cos\tau$, so that condition~(ii) is automatically satisfied.
It can be shown that condition (i) is satisfied as well, e.g., whenever
$\cE(k,N)$ depends polynomially on $k$ and~$N$. More precisely, if $r=\deg\cE$
then $\si^2\sim N^{2r+1}$, while $\si_k^2$ is at most $O(N^{2r})$.

We shall next discuss how the previous developments can be directly applied to two specific spin
chains of HS type, which are of considerable interest in themselves. The first one is the
Polychronakos--Frahm chain of $BC_N$ type~\cite{YT96}, whose
Hamiltonian is given by
\begin{equation*}
  H=\sum_{j\neq k}\Bigg[\frac{1+\ep S_{jk}}{(\xi_j-\xi_k)^2}+\frac{1+\ep S_jS_kS_{jk}}{(\xi_j+\xi_k)^2}\bigg]+\be\sum_k\frac{1-\ep' S_k}{\xi_k^2}\,,
\end{equation*}
where $\ep^2=\ep'^2=1$, $\be>0$, $S_{jk}$ is the operator that permutes the $j$-th and $k$-th
spins and $S_k$ is the operator flipping the $k$-th spin. Moreover, the chain site $\xi_k$ is
expressed in terms of the $k$-th zero $y_k$ of the generalized Laguerre polynomial $L_N^{\be-1}$
as $\xi_k=\sqrt{2y_k}$. It has recently been shown~\cite{BFGR08} that for spin $1/2$ the partition
function of this model is given by
\begin{equation}\label{ZPF}
Z(q)=q^{\frac12N(N-1)\de_{1\ep}}\prod_k\big(1+q^k\big)\,,
\end{equation}
where $\de_{1\ep}$ is the Kronecker delta. Other than the inessential factor
$q^{\frac12N(N-1)\de_{1\ep}}$, which can be removed by shifting the ground state energy, this
partition function is precisely of the form~\eqref{Zk} with $\cE(k,N)=k$.
It follows immediately from our previous discussion that when $N\to\infty$ the spectrum of
$H$ is normally distributed, with mean and variance given by
\[
\mu=\sum_k\frac k2=\frac N4(N+1),\enspace
\si^2=\sum_k\frac{k^2}4=\frac{N}{12}(N+\tfrac12)(N+1).
\]
This fact had been numerically verified in Ref.~\cite{BFGR08}.

The above result has an interesting interpretation in classical partition theory. Indeed,
by Eq.~\eqref{ZPF} the energies of the ferromagnetic chain ($\ep=-1$) are the integers
in the range $0,1,\dots,N(N+1)/2$, the degeneracy of an energy $k$ being the number
$Q_N(k)$ of partitions of the integer $k$ into distinct parts no larger than $N$
(with $Q_N(0)\equiv 1$). We have thus established the asymptotic formula
\[
Q_N(k)\underset{N\to\infty}{\sim}\frac{2^N}{\sqrt{2\pi}\si}\,\e^{-\frac{(k-\mu)^2}{2\si^2}}\,,
\]
(with $k=0,1,\dots,N(N+1)/2$, and $\mu$, $\si$ given in the previous equation) which, to the best
of our knowledge, was not previously known.

The second spin chain we shall consider is the supersymmetric version
of the celebrated Haldane--Shastry chain, with Hamiltonian given by~\cite{BB06}
\begin{equation}\label{H11}
H=\frac12\,\sum_{j<k}\frac{1+\ep P_{jk}}{\sin^2(\vartheta_j-\vartheta_k)}\,,\qquad
\vartheta_k\equiv\frac{k\pi}N\,,
\end{equation}
where $\ep=± 1$. The supersymmetric spin permutation operator $P_{jk}$ acts on an element
$\ket{s_1,\dots,s_N}$ of the spin basis as
  \[
P_{jk}\ket{\dots,s_j,\dots,s_k,\dots}=\ep_{jk}(\bs)\ket{\dots,s_k,\dots,s_j,\dots}\,,
  \]
  where $\ep_{jk}(\bs)$ is $-1$ when either both $s_j$ and $s_k$ are fermionic spins, or $s_j$ and
  $s_k$ are spins of different type with an odd number of fermionic spins between them, and $1$
  otherwise. In the ${\rm su}(1|1)$ case (i.e., when there is only one bosonic and one fermionic
  internal degree of freedom), the partition function of the chain~\eqref{H11} can be written
  as~\cite{BBS08}
\[
Z(q)=2\prod_{k=1}^{N-1}\big(1+q^{k(N-k)}\big)\,,
\]
for both $\ep=±1$. Apart from the irrelevant factor of $2$ (which does not affect the level
density, since it is normalized to $1$), this partition function is a product of $N-1$ factors of
the form~\eqref{Zk} with a polynomial function $\cE(k,N)=k(N-k)$. Hence the above argument
rigorously establishes that in the limit $N\to\infty$ the level density of the
Hamiltonian~\eqref{H11} becomes Gaussian, with parameters
\begin{align*}
&\mu=\frac12\,\sum_kk(N-k)=\frac{N}{12}(N^2-1),\\
&\si^2=\frac14\,\sum_kk^2(N-k)^2=\frac{N}{120}(N^4-1),
\end{align*}
in whole agreement with the numerical computations in Ref.~\cite{BB06}.

To conclude, let us summarize our results and offer some perspectives. We have provided conditions
on the partition function of a finite-dimensional quantum system depending on a positive integer $N$
ensuring that its level density is asymptotically Gaussian as $N\to\infty$. 
Our conditions, which are related to Lyapunov's generalization of the classical central
limit theorem, are directly formulated in terms of the partition function and do not
require the explicit knowledge of the spectrum. We have applied our result to rigorously show
that the level density of two well-known spin chains of HS type becomes Gaussian as the number
of sites tends to infinity. The first chain discussed is
associated with the $BC_N$ root system and presents rational interactions, while the second one
is a supersymmetric version of the original ($A_{N-1}$-type) Haldane--Shastry spin chain.

Our result does not apply to all spin chains of HS type, since the partition function of these
models in general does not factorize as in Eq.~\eqref{Z}. What seems to be true in
this case, however, is that there is an approximate factorization
\begin{equation}\label{Z2}
  Z(q)=\big(1+\ep(N,q)\big)\prod_k Z_k(q) \,,
\end{equation}
where the error term $\ep(N,q)$ and its $q$-derivatives tend to~$0$ in a controlled way as the
number of sites $N$ tends to infinity. Although we shall not further elaborate on this point here,
it is clear that the overall factor $1+\ep(N,q)$ can be taken into account in our approach by
imposing appropriate technical conditions analogous to~\eqref{cond.sigma} and~\eqref{cM}.
We shall provide a more detailed discussion of this issue in a forthcoming paper.

\begin{acknowledgments}
This work was supported in part by the DGI and the Complutense University--CAM
under grants no.~FIS2008-00209 and~GR74/07-910556. A.E.\ acknowledges the financial support by the
Spanish Ministry of Science through a MICINN postdoctoral fellowship.
\end{acknowledgments}
%

\begin{thebibliography}{22}
\expandafter\ifx\csname natexlab\endcsname\relax\def\natexlab#1{#1}\fi
\expandafter\ifx\csname bibnamefont\endcsname\relax
  \def\bibnamefont#1{#1}\fi
\expandafter\ifx\csname bibfnamefont\endcsname\relax
  \def\bibfnamefont#1{#1}\fi
\expandafter\ifx\csname citenamefont\endcsname\relax
  \def\citenamefont#1{#1}\fi
\expandafter\ifx\csname url\endcsname\relax
  \def\url#1{\texttt{#1}}\fi
\expandafter\ifx\csname urlprefix\endcsname\relax\def\urlprefix{URL }\fi
\providecommand{\bibinfo}[2]{#2}
\providecommand{\eprint}[2][]{\url{#2}}

\bibitem[{\citenamefont{Haldane}(1988)}]{Ha88}
\bibinfo{author}{\bibfnamefont{F.~D.~M.} \bibnamefont{Haldane}},
  \bibinfo{journal}{Phys. Rev. Lett.} \textbf{\bibinfo{volume}{60}},
  \bibinfo{pages}{635} (\bibinfo{year}{1988}).

\bibitem[{\citenamefont{Shastry}(1988)}]{Sh88}
\bibinfo{author}{\bibfnamefont{B.~S.} \bibnamefont{Shastry}},
  \bibinfo{journal}{Phys. Rev. Lett.} \textbf{\bibinfo{volume}{60}},
  \bibinfo{pages}{639} (\bibinfo{year}{1988}).

\bibitem[{\citenamefont{Arikawa et~al.}(2001)\citenamefont{Arikawa, Saiga, and
  Kuramoto}}]{ASK01}
\bibinfo{author}{\bibfnamefont{M.}~\bibnamefont{Arikawa}},
  \bibinfo{author}{\bibfnamefont{Y.}~\bibnamefont{Saiga}}, \bibnamefont{and}
  \bibinfo{author}{\bibfnamefont{Y.}~\bibnamefont{Kuramoto}},
  \bibinfo{journal}{Phys. Rev. Lett.} \textbf{\bibinfo{volume}{86}},
  \bibinfo{pages}{3096} (\bibinfo{year}{2001}).

\bibitem[{\citenamefont{Murthy and Shankar}(1994)}]{MS94}
\bibinfo{author}{\bibfnamefont{M.~V.~N.} \bibnamefont{Murthy}}
  \bibnamefont{and} \bibinfo{author}{\bibfnamefont{R.}~\bibnamefont{Shankar}},
  \bibinfo{journal}{Phys. Rev. Lett.} \textbf{\bibinfo{volume}{73}},
  \bibinfo{pages}{3331} (\bibinfo{year}{1994}).

\bibitem[{\citenamefont{Polychronakos}(2006)}]{Po06}
\bibinfo{author}{\bibfnamefont{A.~P.} \bibnamefont{Polychronakos}},
  \bibinfo{journal}{J. Phys. A} \textbf{\bibinfo{volume}{39}},
  \bibinfo{pages}{12793} (\bibinfo{year}{2006}).

\bibitem[{\citenamefont{Hern{á}ndez and L{ó}pez}(2004)}]{HL04}
\bibinfo{author}{\bibfnamefont{R.}~\bibnamefont{Hern{á}ndez}}
  \bibnamefont{and}
  \bibinfo{author}{\bibfnamefont{E.}~\bibnamefont{L{ó}pez}},
  \bibinfo{journal}{JHEP} \textbf{\bibinfo{volume}{0411}}, \bibinfo{pages}{079}
  (\bibinfo{year}{2004}).

\bibitem[{\citenamefont{Enciso et~al.}(2005)\citenamefont{Enciso, Finkel,
  González-López, and Rodr{í}guez}}]{EFGR05}
\bibinfo{author}{\bibfnamefont{A.}~\bibnamefont{Enciso}},
  \bibinfo{author}{\bibfnamefont{F.}~\bibnamefont{Finkel}},
  \bibinfo{author}{\bibfnamefont{A.}~\bibnamefont{González-López}},
  \bibnamefont{and} \bibinfo{author}{\bibfnamefont{M.~A.}
  \bibnamefont{Rodr{í}guez}}, \bibinfo{journal}{Nucl. Phys.}
  \textbf{\bibinfo{volume}{B707}}, \bibinfo{pages}{553} (\bibinfo{year}{2005}).

\bibitem[{\citenamefont{Finkel and González-López}(2005)}]{FG05}
\bibinfo{author}{\bibfnamefont{F.}~\bibnamefont{Finkel}} \bibnamefont{and}
  \bibinfo{author}{\bibfnamefont{A.}~\bibnamefont{González-López}},
  \bibinfo{journal}{Phys. Rev. B} \textbf{\bibinfo{volume}{72}},
  \bibinfo{pages}{174411(6)} (\bibinfo{year}{2005}).

\bibitem[{\citenamefont{Barba et~al.}(2008{\natexlab{a}})\citenamefont{Barba,
  Finkel, González-López, and Rodr{í}guez}}]{BFGR08}
\bibinfo{author}{\bibfnamefont{J.~C.} \bibnamefont{Barba}},
  \bibinfo{author}{\bibfnamefont{F.}~\bibnamefont{Finkel}},
  \bibinfo{author}{\bibfnamefont{A.}~\bibnamefont{González-López}},
  \bibnamefont{and} \bibinfo{author}{\bibfnamefont{M.~A.}
  \bibnamefont{Rodr{í}guez}}, \bibinfo{journal}{Phys. Rev. B}
  \textbf{\bibinfo{volume}{77}}, \bibinfo{pages}{214422(10)}
  (\bibinfo{year}{2008}{\natexlab{a}}).

\bibitem[{\citenamefont{Barba et~al.}(2008{\natexlab{b}})\citenamefont{Barba,
  Finkel, González-López, and Rodr{í}guez}}]{BFGR08epl}
\bibinfo{author}{\bibfnamefont{J.~C.} \bibnamefont{Barba}},
  \bibinfo{author}{\bibfnamefont{F.}~\bibnamefont{Finkel}},
  \bibinfo{author}{\bibfnamefont{A.}~\bibnamefont{González-López}},
  \bibnamefont{and} \bibinfo{author}{\bibfnamefont{M.~A.}
  \bibnamefont{Rodr{í}guez}}, \bibinfo{journal}{Europhys. Lett.}
  \textbf{\bibinfo{volume}{83}}, \bibinfo{pages}{27005(6)}
  (\bibinfo{year}{2008}{\natexlab{b}}).

\bibitem[{\citenamefont{Basu-Mallick and Bondyopadhaya}(2006)}]{BB06}
\bibinfo{author}{\bibfnamefont{B.}~\bibnamefont{Basu-Mallick}}
  \bibnamefont{and}
  \bibinfo{author}{\bibfnamefont{N.}~\bibnamefont{Bondyopadhaya}},
  \bibinfo{journal}{Nucl. Phys.} \textbf{\bibinfo{volume}{B757}},
  \bibinfo{pages}{280} (\bibinfo{year}{2006}).

\bibitem[{\citenamefont{Basu-Mallick et~al.}(2009)\citenamefont{Basu-Mallick,
  Finkel, and González-López}}]{BFG09}
\bibinfo{author}{\bibfnamefont{B.}~\bibnamefont{Basu-Mallick}},
  \bibinfo{author}{\bibfnamefont{F.}~\bibnamefont{Finkel}}, \bibnamefont{and}
  \bibinfo{author}{\bibfnamefont{A.}~\bibnamefont{González-López}},
  \bibinfo{journal}{Nucl. Phys.} \textbf{\bibinfo{volume}{B812}},
  \bibinfo{pages}{402} (\bibinfo{year}{2009}).

\bibitem[{\citenamefont{Basu-Mallick and Bondyopadhaya}()}]{BB09}
\bibinfo{author}{\bibfnamefont{B.}~\bibnamefont{Basu-Mallick}}
  \bibnamefont{and}
  \bibinfo{author}{\bibfnamefont{N.}~\bibnamefont{Bondyopadhaya}},
  \bibinfo{note}{arXiv:0811.3110v1 [cond-mat.stat-mech]}.

\bibitem[{\citenamefont{Haake}(2001)}]{Ha01}
\bibinfo{author}{\bibfnamefont{F.}~\bibnamefont{Haake}},
  \emph{\bibinfo{title}{{Q}uantum {S}ignatures of {C}haos}}
  (\bibinfo{publisher}{Springer-Verlag}, \bibinfo{address}{Berlin},
  \bibinfo{year}{2001}), \bibinfo{edition}{2nd} ed.

\bibitem[{\citenamefont{Berry and Tabor}(1977)}]{BT77}
\bibinfo{author}{\bibfnamefont{M.~V.} \bibnamefont{Berry}} \bibnamefont{and}
  \bibinfo{author}{\bibfnamefont{M.}~\bibnamefont{Tabor}},
  \bibinfo{journal}{Proc. R. Soc. Lond. A} \textbf{\bibinfo{volume}{356}},
  \bibinfo{pages}{375} (\bibinfo{year}{1977}).

\bibitem[{\citenamefont{Mehta}(2004)}]{Me04}
\bibinfo{author}{\bibfnamefont{M.~L.} \bibnamefont{Mehta}},
  \emph{\bibinfo{title}{{R}andom {M}atrices}} (\bibinfo{publisher}{Elsevier},
  \bibinfo{address}{San Diego}, \bibinfo{year}{2004}), \bibinfo{edition}{3rd}
  ed.

\bibitem[{\citenamefont{Polychronakos}(1993)}]{Po93}
\bibinfo{author}{\bibfnamefont{A.~P.} \bibnamefont{Polychronakos}},
  \bibinfo{journal}{Phys. Rev. Lett.} \textbf{\bibinfo{volume}{70}},
  \bibinfo{pages}{2329} (\bibinfo{year}{1993}).

\bibitem[{\citenamefont{Frahm}(1993)}]{Fr93}
\bibinfo{author}{\bibfnamefont{H.}~\bibnamefont{Frahm}}, \bibinfo{journal}{J.
  Phys. A} \textbf{\bibinfo{volume}{26}}, \bibinfo{pages}{L473}
  (\bibinfo{year}{1993}).

\bibitem[{\citenamefont{Polychronakos}(1994)}]{Po94}
\bibinfo{author}{\bibfnamefont{A.~P.} \bibnamefont{Polychronakos}},
  \bibinfo{journal}{Nucl. Phys.} \textbf{\bibinfo{volume}{B419}},
  \bibinfo{pages}{553} (\bibinfo{year}{1994}).

\bibitem[{\citenamefont{Ash and Dol{é}ans-Dade}(2000)}]{AD00}
\bibinfo{author}{\bibfnamefont{R.~B.} \bibnamefont{Ash}} \bibnamefont{and}
  \bibinfo{author}{\bibfnamefont{C.~A.} \bibnamefont{Dol{é}ans-Dade}},
  \emph{\bibinfo{title}{Probability and Measure Theory}}
  (\bibinfo{publisher}{Academic Press}, \bibinfo{address}{San Diego},
  \bibinfo{year}{2000}), \bibinfo{edition}{2nd} ed.

\bibitem[{\citenamefont{Yamamoto and Tsuchiya}(1996)}]{YT96}
\bibinfo{author}{\bibfnamefont{T.}~\bibnamefont{Yamamoto}} \bibnamefont{and}
  \bibinfo{author}{\bibfnamefont{O.}~\bibnamefont{Tsuchiya}},
  \bibinfo{journal}{J. Phys. A} \textbf{\bibinfo{volume}{29}},
  \bibinfo{pages}{3977} (\bibinfo{year}{1996}).

\bibitem[{\citenamefont{Basu-Mallick et~al.}(2008)\citenamefont{Basu-Mallick,
  Bondyopadhaya, and Sen}}]{BBS08}
\bibinfo{author}{\bibfnamefont{B.}~\bibnamefont{Basu-Mallick}},
  \bibinfo{author}{\bibfnamefont{N.}~\bibnamefont{Bondyopadhaya}},
  \bibnamefont{and} \bibinfo{author}{\bibfnamefont{D.}~\bibnamefont{Sen}},
  \bibinfo{journal}{Nucl. Phys.} \textbf{\bibinfo{volume}{B795}},
  \bibinfo{pages}{596} (\bibinfo{year}{2008}).

\end{thebibliography}

\end{document}